\documentclass[aps,prl,preprintnumbers,showpacs,nofootinbib,floatfix]{revtex4}
\usepackage{graphicx}
\usepackage{comment}
\usepackage{dcolumn}
\usepackage{bm}
\begin{document}
\newcommand{\newc}{\newcommand}
\newc{\ra}{\rightarrow}
\newc{\lra}{\leftrightarrow}
\newc{\beq}{\begin{equation}}
\newc{\eeq}{\end{equation}}
\newc{\barr}{\begin{eqnarray}}
\newc{\earr}{\end{eqnarray}}
\newcommand{\Od}{{\cal O}}
\newcommand{\lsim}   {\mathrel{\mathop{\kern 0pt \rlap
  {\raise.2ex\hbox{$<$}}}
  \lower.9ex\hbox{\kern-.190em $\sim$}}}
\newcommand{\gsim}   {\mathrel{\mathop{\kern 0pt \rlap
  {\raise.2ex\hbox{$>$}}}
  \lower.9ex\hbox{\kern-.190em $\sim$}}}
\title{PIONIC CONTRIBUTION TO NEUTRINOLESS DOUBLE BETA DECAY}

\author{J. D. Vergados$^{(1),(2)}$\thanks{Vergados@uoi.gr},  Amand Faessler$^{(3)}$,  and  H. Toki$^{(4)}$}
\affiliation{$^{(1)}${\it Physics Department, University of Ioannina, Ioannina, GR 451 10, Greece,}}
\affiliation{$^{(2)}${\it  Theory Division, CERN, Geneva,Switzwerland}}
\affiliation{$^{(3)}${\it Institute f$\ddot {u}$r Theoterische Physik,Universit$\ddot{a}$t T$\ddot{u}$bingen, Germany,}}
\affiliation{$^{(4)}${\it RCNP, Osaka University, Osaka, 567-0047, Japan}}

\begin{abstract}
It is well known that neutrinoless double decay is going to play a
crucial role in settling the neutrino properties, which cannot be
extracted from the neutrino oscillation data. It is, in
particular, expected to settle the absolute scale of neutrino mass
and determine whether the neutrinos are Majorana particles, i.e.
they coincide with their own antiparticles. In order to extract
the average neutrino mass from the data one must be able to
estimate the contribution all  possible high mass intermediate particles. The latter, which
occur in practically all extensions of the standard model, can, in principle,
be differentiated from the usual mass term, if data from various targets are available. One, however,
must first be able reliably calculate the corresponding nuclear matrix elements.
Such calculations are extremely
difficult since the effective transition operators are very short
ranged. For such operators processes like  pionic contributions,
which are usually negligible, turn out to be dominant. We study
such an effect in a non relativistic quark model for the pion and the nucleon.

\end{abstract}

\date{\today}
\maketitle
\section{Introduction}
The discovery of neutrino oscillations can be considered as one of the greatest triumphs of modern physics.
It began with atmospheric neutrino oscillations \cite{SUPERKAMIOKANDE}interpreted as
 $\nu_{\mu} \rightarrow \nu_{\tau}$ oscillations, as well as
 $\nu_e$ disappearance in solar neutrinos \cite{SOLAROSC}. These
 results have been recently confirmed by the KamLAND experiment \cite{KAMLAND},
 which exhibits evidence for reactor antineutrino disappearance.
  As a result of these experiments we have a pretty good idea of the neutrino
mixing matrix and of the two independent quantities $\Delta m^2$, e.g $m_2^2-m^2_1$ and $m^2_3-m^2_2$.
 Fortunately these
two  $\Delta m^2$ values are vastly different, $$|\Delta
m^2_{21}|=|m_2^2-m_1^2|=(5.0-7.5)\times 10^{-5}(eV)^2$$ and
$$|\Delta m^2_{32}|=|m_3^2-m_2^2|=2.5\times 10^{-3}(eV)^2.$$
 This means that the relevant $L/E$ parameters are very different. Thus for a given energy the experimental
 results can approximately be described as two generation oscillations. For an accurate description  of the data,
 however, a three generation analysis  \cite{BAHCALL02}-\cite{BARGER02} is necessary.

 We thus know that the neutrinos are massive, with two non zero $\Delta m^2$, and they are admixed.
 We do not know, however, whether they are
 Majorana, i.e. the mass eigenstates coincide with their antiparticles, or of Dirac type, i.e. the mass
 eigenstates do not coincide with their antiparicles. Furthermore we do not know the absolute mass scale as well as the sign of $\Delta m^2_{32}$. The first question can be settled by neutrinoless double beta decay ($0\nu~\beta\beta-$ decay). The second will also, most likely,  be settled by this  process.
 
  We should stress, of course, the fact that the light neutrino mediated process is not the only mechanism available for $0\nu\beta\beta$ \cite{JDV02}. Among those are some which involve heavy intermediate particles. These lead to very short ranged two body effective transition operators, which must be dealt with care, due to the presence of the
nuclear hard core. To this end three treatments have been proposed:
\begin{itemize}
\item Treat the nucleons as composite particles (two nucleon mode).\\
This can be done in the context of non relativistic quark model or simply by assigning to the nucleon a
suitable form factor \cite{JDV1981}.
\item Consider the possibility of six quark cluster in the nucleus \cite{JDV85}
\item Consider other particles in the nuclear soup.\\
 The most prominent are pions in flight between the two interacting nucleons \cite{JDV02}
\end{itemize}
In the present study we will examine the last possibility. This was examined long time ago \cite{JDV87}
and it was revived in the context of R-parity violating supersymmetry a decade later \cite{FKSW97,FKS98,WKS99}
as well as recently\cite{FGKS07}.
It was
shown that in the context of R-parity violating supersymmetry the pion mode is more important than the two nucleon mechanism. The same conclusion
was reached recently in the context of effective field theory\cite{PR-MV03}.\\
In the above  treatments the pions
were treated as elementary particles. This approach is reasonable in particle physics, but one knows, of course, that the hadrons involved are not elementary. Furthermore a crucial factorization approximation has to be made, by inserting only the vacuum as intermediate state, (see Eqs (\ref{Eq:EP2pi}) and (\ref{Eq:EP1pi}) below). Finally, even though the hadrons are elementary, in the interesting  case of the pseudoscalar coupling an assumption had to be made about the quark mass, taken to be the current quark  mass.\\
 In this work we are going to adopt a different procedure. The hadrons will be assumed to have a quark substructure in the context of the harmonic oscillator. In the harmonic oscillator approximation the internal degrees of freedom can be
 separated from the center of mass motion.  In this approach one derives the effective operator at the quark level by  a suitable non relativistic expansion of the elementary amplitude. In some processes in our formalism one extra $q{\bar q}$ pair must be produced. This can can be achieved either through the weak interaction itself or via the strong interaction. The net result is that, in
this new approach, one obtains  new types of operators, including some
 that are non local at the nucleon level. One must weigh these advantages, however, against possible shortcomings of the need for a non relativistic reduction of the transition operator  at the quark level. 
\section{The contribution of pions in flight between nucleons}
\begin{figure}[!ht]
 \begin{center}
\includegraphics[scale=0.6]{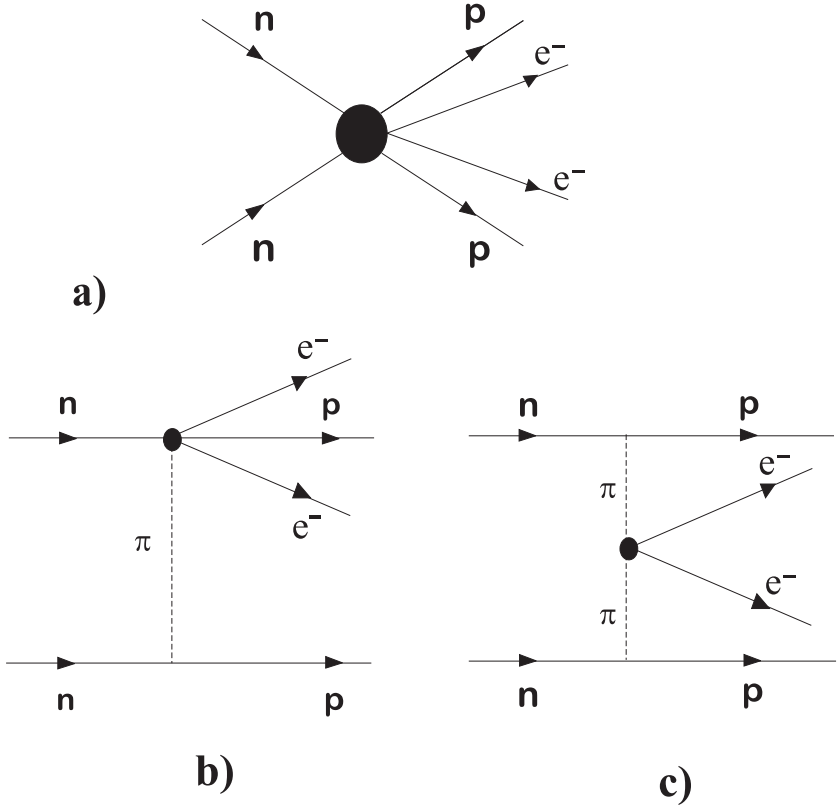}
 \caption{The double beta decay of two neutrons into two protons  at the two  nucleon level (a) arising when
all the intermediate particles at the quark level are very heavy. The double beta decay of a neutron with the simultaneous production of a $\pi^+$, which is then absorbed by another neutron converting it into a proton (b)
(one pion mode). A neutron can also be converted into a proton and a  $\pi^-$  . The $\pi^-$ then
double beta decays into a $\pi^+$,
which subsequently is absorbed by another neutron converting into a proton (c) (two pion mode).
    }
    \label{fig:pions}
   \end{center}
  \end{figure}

As we have mentioned in the introduction when the intermediate fermion, e.g. the Majorana neutrino, is very
heavy the transition operator becomes very short ranged. In this case the usual two nucleon mechanism may
be suppressed due to the nuclear hard core
and the contribution of other particles in the nuclear soup, such as pions, may dominate. These mechanisms at the nucleon level are
illustrated in Fig. \ref{fig:pions}.

The two body double beta decay operator, associated with heavy
intermediate particle exchange, will be normalized in a way which
is consistent with the light intermediate neutrino. We begin with the intermediate heavy neutrino. Then:
\beq \eta_{\nu}\frac{R_0}{r}\Leftrightarrow \eta^{L,R}_N \frac{4 \pi
R_0}{m_e m_p}\delta({\bf r}_1-{\bf r}_2) \label{op1}
 \eeq 
 $L,R$ stand for leftt handed and right habded currents respectively with
\beq \eta_{\nu}=\frac{\prec m_{\nu}\succ }{m_e}~,~\eta^{L,R}_{N}=\prec
\frac{m_p}{ m_{N}} \succ \label{op3}
\eeq
The corresponding
expression in momentum space becomes:
 \beq
\eta_{\nu}\frac{R_0}{r}\delta({\bf r}_1-{\bf r}'_1)\delta({\bf
r}_2-{\bf r}'_2)\Leftrightarrow \eta^{L,R}_N \frac{4 \pi R_0}{m_e
m_p}\Omega_{\beta \beta}
\eeq
\beq
\Omega_{\beta \beta}=\frac{1}{(2 \pi )^3} \delta({\bf p}_1+{\bf p}_2-{\bf
p}'_1-{\bf p}'_2) A({\bf p}_1,{\bf p}_2,{\bf p}'_1, {\bf p}'_2)
\label{op4}
\eeq
The function $ A({\bf p}_1,{\bf p}_2,{\bf p}'_1, {\bf p}'_2)$ depends on the assumed mechanism for the neutrinoless double beta decay.

 The factor $\eta^{L,R}_N$ is not usually
included in the nuclear matrix element. The factor
$\frac{R_0m_p}{m_e}$
 will be absorbed into the effective nuclear operator, while the
factor $\frac{4 \pi}{m^2_p}$ will eventually be included in the
effective coupling, as will be discussed
 in this work.

 With  the above expressions the formula for the life time due to heavy intermediate neutrinos in left handed V-A theories can be cast in
  the form:
 \beq
 [T_{1/2}^{0\nu}]^{-1} = G_{01}\left[\eta^L_N \left( (\frac{f_V}{f_A})^2 \Omega_F-\Omega_{GT}+\alpha_{1\pi} \Omega_{1\pi} +\alpha_{2\pi} \Omega_{2\pi} \right) \right ]^2
 \eeq
 The two nucleon  contribution $ (\frac{f_V}{f_A})^2\Omega_F-\Omega_{GT}$ was inserted in the above
equation merely for comparison.

 The case of other heavy intermediate particles, as those encountered in the R-parity violating supersymmetry can be handled in a similar fashion:
\beq
[T_{1/2}^{0\nu}]^{-1} = G_{01}\left [
\frac{3}{8}(\eta^T+\frac{5}{3}\eta^{PS})\left(\frac{4}{3} \alpha_{1\pi}\Omega_{1\pi}+\alpha_{2\pi}\Omega_{2\pi} \right) \right ]^2
 \eeq
    \beq
    \Omega_{k\pi}~ = \frac{m_p}{m_e}~
        [ M^{k\pi}_{GT}+M^{k\pi}_T].
    \label{eq:6.6}
    \eeq
    
    In both cases:
    \beq
    M^{k\pi}_{GT}=\sum_{i<j}\tau_{+}(i)\tau_{+}(j)\sigma_i.\sigma_j~\frac{R_0}{r} F^{(k)}_1(x_{\pi})
    \eeq
    \beq
    M^{k\pi}_{T}=\sum_{i<j}\tau_{+}(i)\tau_{+}(j)\left[3\sigma_i.\hat{r}_{ij}\sigma_j.\hat{r}_{ij}-\sigma_i.\sigma_j \right]~\frac{R_0}{r} F^{(k)}_2(x_{\pi})
    \eeq
    Where $R_0$ is the nuclear radius, $x_{\pi}=m_{\pi} r_{ij}$ and
    \beq
    F^{(1)}_1(x)=e^{-x},F^{(1)}_2(x)=(x^2+3 x+3)e^{-x},F^{(2)}_1(x)=(x-2)e^{-x},F^{(2)}_2(x)=(x+1)e^{-x}
    \label{Eq:Frs}
    \eeq
    
The function $  A({\bf p}_1,{\bf p}_2,{\bf p}'_1, {\bf p}'_2)$ depends on the pion mode  under consideration.
\section{The 2-pion mode}
\begin{figure}[!ht]
 \begin{center}
\includegraphics[scale=1.2]{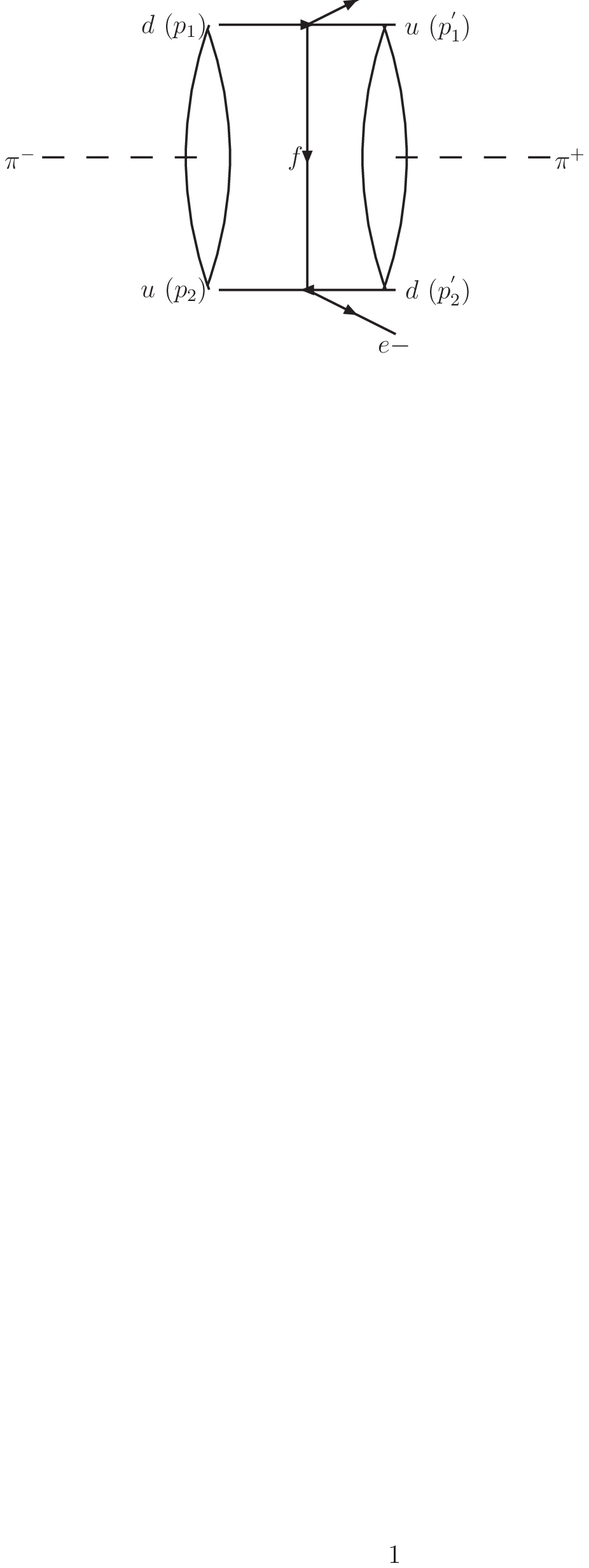}
 \caption{The $0\nu \beta\beta$ decay of pions in flight (2$\pi$ mode of Fig. \ref{fig:pions}) illustrated at the quark level.
 $f$ stands for a effective exchange of a heavy Majorana fermion ( heavy neutrino or, as in
 R-parity violating supersymmetry, a neutralino, gluino etc). The ellipses merely indicate that the pion is a bound state of two quarks.}
    \label{fig:2pions}
   \end{center}
  \end{figure}

The spin dependence of the transition operator is in this case trivial. So we will focus on the orbital structure of  of the operator 
The function $ A({\bf p}_1,{\bf p}_2,{\bf p}'_1, {\bf p}'_2)$  is independent of the momenta in the standard V-A theory as well as in the case of the scalar (S-S) theory. It is, however,  a model dependent function in the case of psedoscalar (P-P) interaction encountered, e.g.,  in R-Parity violating SUSY mediated double beta decay. In the last case we find
\beq
A=-\frac{1}{3}{\bf A}_1.{\bf A}_2
\eeq
where ${\bf A}_i$ is the amplitude resulting from the non relativistic reduction of the pseudoscalar
involved in the $d\rightarrow u$ coupling, i.e.
\beq
\bar{u}(p'_i)\gamma_5 d(p_i) \rightarrow  {\bf A}_i. \mbox{ \boldmath $\sigma_i$}
\eeq
where \mbox{ \boldmath $\sigma_i$} is the spin of the quark $i$ and
\beq
A_i=\frac{1}{2m_d}{\bf p}_i-\frac{1}{2m_u}{\bf p}'_i
\eeq
We find it convenient to rewrite them as follows:
\beq
{\bf A}_1=-\frac{1}{\sqrt{2}}(\frac{\mbox{ \boldmath $\rho$}}{2m_d}-\frac{\mbox{ \boldmath $\rho$'}}{2m_u})
+(\frac{1}{2m_d}-\frac{1}{2m_u})\frac{{\bf q}}{2}
\eeq
\beq
{\bf A}_2=\frac{1}{\sqrt{2}}(\frac{\mbox{ \boldmath $\rho$}}{2m_d}-\frac{\mbox{ \boldmath $\rho$'}}{2m_u})
+(\frac{1}{2m_d}-\frac{1}{2m_u})\frac{{\bf q}}{2}
\eeq
Where ${\bf q=P}_{\pi}$ is the momentum of the pion in flight between the two nucleons 
and  $\mbox{ \boldmath $\rho$}$ and $\mbox{ \boldmath $\rho^{'}$}$ are the relative internal momenta (see next subsection). One normally ignores at this level the
momentum carried away by the two leptons.
    The $2 \pi$  $0 \nu-\beta \beta$ decay contribution in the case of heavy Majorana neutrino or any other
Majorana fermion is  explicitly shown
 in Fig. \ref{fig:2pions}.
  \subsection{Orbital integrals in the two pion exchange.}
  The pion wave function is given by:
  \beq
\psi_{{\bf P}_{\pi}}({\bf Q},\rho)= \sqrt{2 E_{\pi}}\left (2 \sqrt{2} \right )^{1/2} \left ( 2 \pi \right ) ^{3/2} \delta \left ( \sqrt{2}{ \bf Q}-{\bf P}_{\pi} \right ) \phi_{\pi}( \mbox{ \boldmath $\rho$} )
\eeq
where ${\bf P}_{\pi}$ is the pion momentum and
\beq
 \mbox{ \boldmath $\rho$}=\frac{1}{\sqrt{2}}\left ({\bf p}_2-{\bf p}_1 \right)~,~{\bf Q}=\frac{1}{\sqrt{2}}\left ({\bf p}_2+{\bf p}_1 \right )
\eeq
 with
${\bf p}_1$ and ${\bf p}_2$ being the momenta of the quark and antiquark participating  in the pion.
This wave function is normalized in the usual way:
 \beq
 \prec \psi_{{\bf P}_{\pi}}|\psi_{{\bf P}'_{\pi}}\succ = 2E_{\pi} (2 \pi)^3 \delta({\bf P}_{\pi}-{\bf P}'_{\pi})
\label{wfpion}
\eeq
$\phi_{\pi}( \mbox{ \boldmath $\rho$} )$ is described by an $1s$ harmonic oscillator state. In momentum space it takes the form:
\beq
\phi_{\pi}( \mbox{ \boldmath $\rho$})=\phi_{\pi}(0) e^{-(b^2_{\pi}\rho^2)/2},
~~\phi_{\pi}(0)=\sqrt{\frac{b^3_{\pi}}{\pi \sqrt{\pi}}}
\label{Eq:pionwf}
\eeq
Thus the orbital matrix element in this case  takes
the form: \beq ME_{2 \pi}={\cal M}_{2 \pi}(2\pi )^3  \delta ({\bf
P}_{\pi}-{\bf P}'_{\pi})~,~ {\cal M}_{2 \pi}=\frac{1}{2 \pi
\sqrt{2 \pi}} \frac{2 m_{\pi}}{b^3_N}f^{(1)}_{2\pi}(x),~~f^{(1)}_{2\pi}(x)=\frac{1}{x^3}
\eeq
where $x=\frac{b_{\pi}}{b_N}$.  $b_{\pi}$ and $b_N$ are the
harmonic oscillator (HO) size parameters for the pion and the
nucleon respectively. We have decided to introduce the ratio $x$
as a variable to be adjusted.\\
In V-A theories after incorporating the spin we find:
 \beq \frac{4
\pi}{m^2_p} {\cal M}_{2 \pi} = c_{2 \pi} m^2_{\pi} \eeq with \beq
c_{2 \pi}=\frac{1}{ \sqrt{2 \pi}} \frac{4}{b^3_N m_p^2
m_{\pi}}
\prec |1- \mbox{\boldmath $\sigma_1$}.\mbox{\boldmath
$\sigma_2$}|\succ f^{(1)}_{2\pi}(x)
 \eeq

  where $\prec | \mbox{\boldmath
$\sigma_1$}.\mbox{\boldmath $\sigma_2$}|\succ=-3$ is the spin ME.
 One now can  construct the effective transition operator
in coordinate space at the nuclear level. The effective coupling
in V-A  theory is given \cite{JDV02} by:
 \beq
  \alpha_{2 \pi} = c_{2 \pi} g_r^2 \left (\frac{m_{\pi}}{2m_N}
\right )^2 \frac{1}{4 \pi} \frac{1}{6  m^2_{\pi}} \frac{1}{f_A^2}
 \eeq
 Or
 \beq
  \alpha_{2 \pi}=\frac{2}{3 f^2_A}  f^2_{\pi N N} c_{2\pi}
\label{alpha2pi}
  \eeq
  Using $f^2_{\pi N N}=0.08 $ and $b_N=1.0$ fm we find $\alpha_{2 \pi}=0.013$ and $0.11$ for $x=1.0$
and $0.5$ respectively. For the scalar interaction one gets the value $f_S^2/4$ with the value of $f_S$ depending
on the specific particle model.

\begin{figure}[!ht]
 \begin{center}
\rotatebox{90}{\hspace{-0.0cm} {$f^{(1)}_{2\pi}(x) \longrightarrow$}}
\includegraphics[scale=0.6]{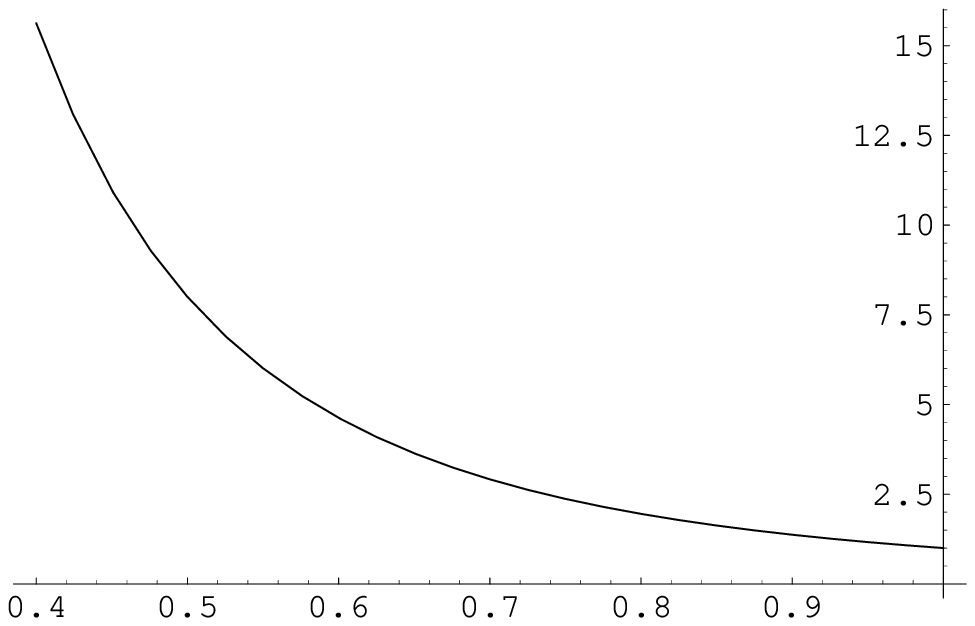}
\rotatebox{90}{\hspace{-0.0cm} {$f^{(2)}_{2\pi}(x) \longrightarrow$}}
\includegraphics[scale=0.6]{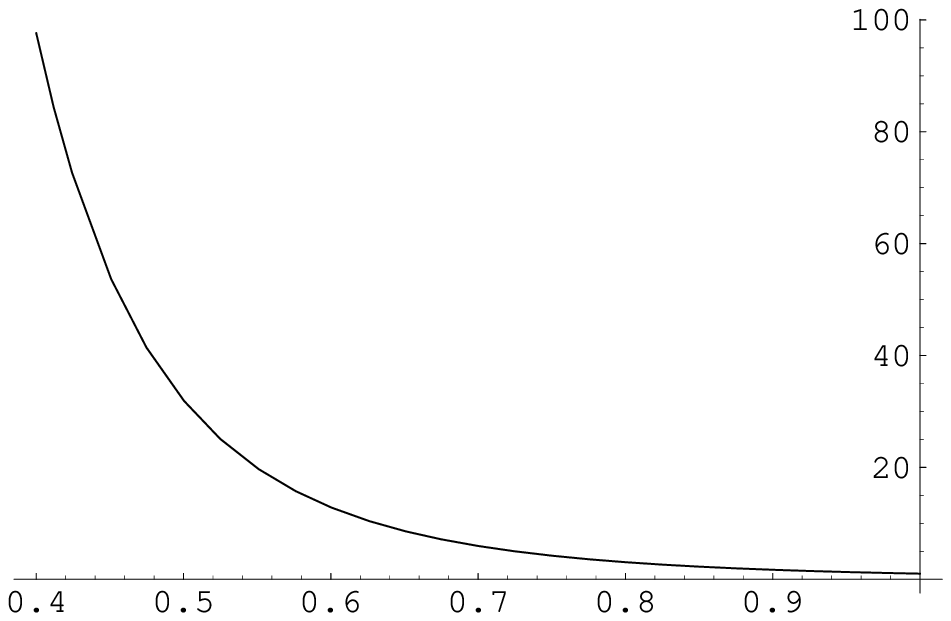}\\
{\hspace{-2.0cm}} {\hspace{-2.0cm}} {\hspace{4.0cm}
$\frac{b_{\pi}}{b_N}\longrightarrow$}
 \caption{ The function $f^{(1)}_{2\pi}(x)$ on the left and $ f^{(2)}_{2\pi}(x)$ on the right as a function of $x=\frac{b_{\pi}}{b_N}$.
}
  \label{fig:2piVA}
   \end{center}
  \end{figure}

The dependence of the results on the pion size parameter is
exhibited in Figs \ref{fig:2piVA}.\\
  In the case
of the pseudoscalar coupling, since the pion has spin zero, we
encounter the combination:
 \beq
 ({\bf A}_1. \mbox{ \boldmath $\sigma_1$})({\bf A}_2. \mbox{ \boldmath $\sigma_2$})\Rightarrow
 -\frac{1}{3}({\bf A}_1.{\bf A}_2) (\mbox{ \boldmath $\sigma_1$}.\mbox{ \boldmath $\sigma_2$})
 \eeq
In this case one can show that  the orbital amplitude is
\beq {\cal M}_{2 \pi}=\frac{1}{2
\pi \sqrt{2 \pi}} \frac{2 m_{\pi}}{b^3_N} \left(
\frac{1}{4}(\kappa^2_d+\kappa^2_u )
 -\frac{1}{6}(\kappa_d -\kappa_u )^2  b_N^2 {\bf q}^2 \right )f^{(2)}_{2\pi}(x),~~f^{(2)}_{2\pi}(x)=\frac{1}{x^5}
 \eeq
 Where $ {\bf q}$ the momentum of the propagating pion and
 \beq
 \kappa_d=\frac{1}{2 m_d b_N}~,~\kappa_u=\frac{1}{2 m_u b_N}
 \eeq
 The above equation can be rewritten in a way that the pion propagator is manifest:
 \barr {\cal M}_{2 \pi}&=&\frac{1}{2
 \pi \sqrt{2 \pi}} \frac{2 m_{\pi}}{b^3_N}
\nonumber\\
&&\left(
\frac{1}{4}(\kappa^2_d+\kappa^2_u )+\frac{1}{6}(\kappa_d -\kappa_u )^2  b^2_N m^2_{\pi}
 -\frac{1}{6}(\kappa_d -\kappa_u )^2  b_N^2 ({\bf q}^2 +m^2_{\pi}) \right )f^{(2)}_{2\pi}(x)
 \earr
 In other words  there appear two terms $ c^0_{
2 \pi}$ and $c^q_{2 \pi}b^2_N (q^2+m^2_{\pi})$ with
 \beq
 c^0_{2 \pi}=\frac{1}{ \sqrt{2 \pi}} \frac{4}{b^3_N m_p^2
m_{\pi}} f^{(2)}_{2\pi}(x)\left ( \frac{1}{4}( \kappa^2_d+\kappa^2_u )+
\frac{1}{6}(\kappa_d -\kappa_u )^2  b^2_N m^2_{\pi} \right )
  \eeq
 \beq
 c^q_{2 \pi}=-\frac{1}{ \sqrt{2 \pi}} \frac{4}{b^3_N m_p^2
m_{\pi}}\frac{1}{6}(\kappa_d -\kappa_u )^2 b^2_N m^2_{\pi} f^{(2)}_{2\pi}(x)
  \eeq
The first gives rise to an effective operator similar to that of
the V-A theory with a coupling
 \beq
  \alpha_{2 \pi}=\frac{2}{3 f^2_A} \frac{1}{\sqrt{2 \pi}} f^2_{\pi N N}
   \frac{1}{m_{\pi} m_p^2 b^3_N}f^{(2)}_{2\pi}(x) \left (\frac{1}{4}( \kappa^2_d+\kappa^2_u )
+\frac{1}{6}(\kappa_d -\kappa_u )^2  b^2_N m^2_{\pi} \right )
   \prec | \mbox{\boldmath
$\sigma_1$}.\mbox{\boldmath $\sigma_2$}|\succ
  \eeq
  with $\prec | \mbox{\boldmath
$\sigma_1$}.\mbox{\boldmath $\sigma_2$}|\succ=-3.$

  The second term, contributing when the u and d quarks are not degenerate,
   yields a  coupling $ \alpha_{2 \pi}(\Omega_{1 \pi})$   where:
  \beq
  \alpha_{2 \pi}(\Omega_{1 \pi}) =-\frac{4}{ f^2_A} \frac{1}{\sqrt{2 \pi}}
  f^2_{\pi N N} \frac{m_{\pi}}{m_p} \frac{1}{ m_p b_N} f^{(2)}_{2\pi}(x)
   \frac{1}{6}(\kappa_d -\kappa_u )^2 )
   \prec | \mbox{\boldmath
$\sigma_1$}.\mbox{\boldmath $\sigma_2$}|\succ
  \eeq
  which is associated with the operator with one pion propagator less, i.e. that encountered in the $1 \pi$ mode (see below). Such an operator is absent in the elementary particle treatment, even though the quarks are assumed to be non degenarate.

\section{The 1-pion mode}
\begin{figure}[!ht]
 \begin{center}
\includegraphics[scale=1.2]{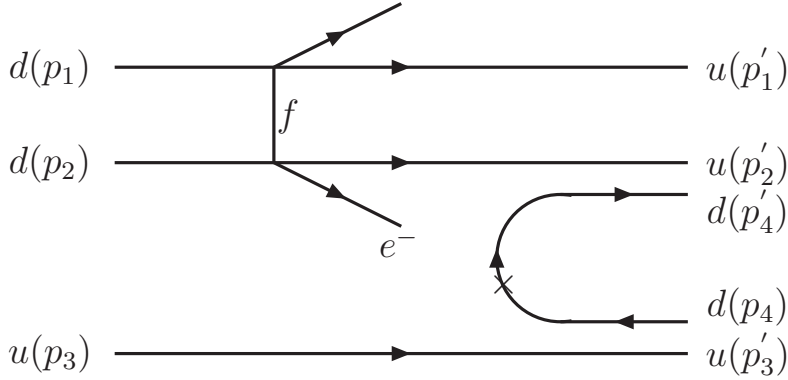}\\
 \caption{  The pion mediated $0\nu~\beta\beta$ decay in the so-called  $1\pi$ mode. At the top we show  the diagram in which the quarks of the pion are spectators
, i.e. the  heavy intermediate heavy fermion f is exchanged between the other two quarks. $\times $ indicates that a $q{\bar q}$ pair is created out of the vacuum in the context of a multigluon exchange.
We will call it direct diagram.}
    \label{fig:pionb}
   \end{center}
  \end{figure}

\begin{figure}[!ht]
 \begin{center}
\includegraphics[scale=1.2]{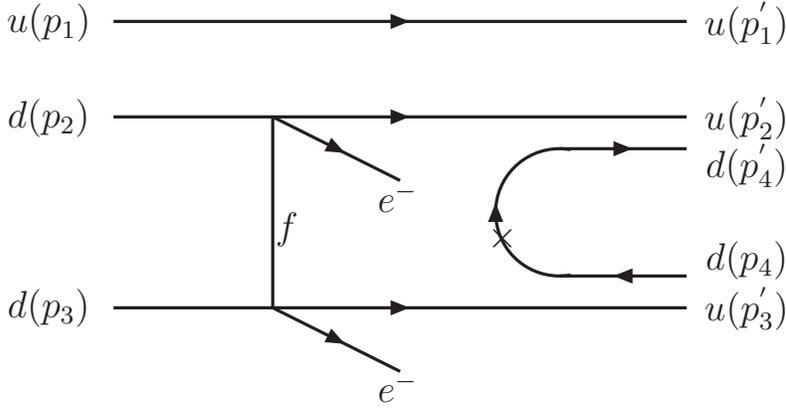}
 \caption{  The same as in Fig. \ref{fig:pionb} involving  the  exchange diagram. In this case the quark involved in the pion participates
in  the exchange of the heavy   fermion f , co-operating this way with another quark belonging in
 the nucleon.}
    \label{fig:pionc}
   \end{center}
  \end{figure}

\begin{figure}[!ht]
 \begin{center}
\includegraphics[scale=1.2]{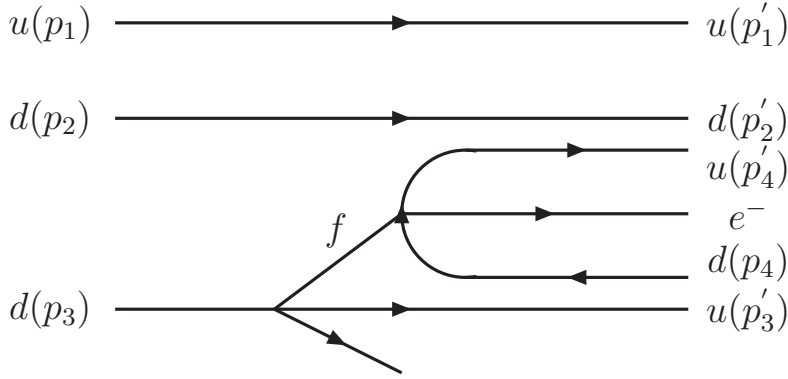}
 \caption{  The same as in Fig. \ref{fig:pionc} but in  a novel mechanism, i.e. one in which the $q{\bar q}$
pair is produced by the weak  interaction itself. }
    \label{fig:piond}
   \end{center}
  \end{figure}

In this case a positively charged pion, produced in virtual double beta decay of a neutron into a proton,  is absorbed by another neutron converting it
into a proton. At the quark level the first of these steps  is exhibited in
Figs \ref{fig:pionb}-\ref{fig:piond}. In these figures   a $q\bar{q}$ pair is created out of the vacuum. In the first two figures this is achieved as, e.g., in a gluon exchange \cite{HOV86} or a multigluon exchange simulated   in the $^3P_0$ model \cite{Micu},\cite{LOPR73},\cite{LOPR74},\cite{LOPR77}. The latter is a fairly old model, which still continues to be successfully applied in the description of meson decays \cite{SSQH08}. In Fig. \ref{fig:piond} this pair is created by the weak interaction itself.

\subsection{The orbital part at the quark level}
Orbital wave functions in momentum space are expressed in terms of Jacobi coordinates:
\beq
\psi_{{\bf P}_{\pi}}= \sqrt{2 E_{\pi}}\left (2 \sqrt{2} \right )^{1/2} \left ( 2 \pi \right ) ^{3/2} \delta \left ( \sqrt{2}{ \bf Q}_{\pi}-{\bf P}_{\pi} \right ) \phi_{\pi}( \mbox{ \boldmath $\rho$} )
\eeq
\beq
\psi_{{\bf P}}= \left (3 \sqrt{3} \right )^{1/2} \left ( 2 \pi \right ) ^{3/2} \delta \left ( \sqrt{3}{ \bf Q-P}_{} \right ) \phi(\mbox{ \boldmath $\xi$},\mbox{ \boldmath $\eta$})
\eeq
\beq
\psi_{{\bf P'}}= \left (3 \sqrt{3} \right )^{1/2} \left ( 2 \pi \right ) ^{3/2} \delta \left ( \sqrt{3}{ \bf Q'-P'}_{} \right ) \phi(\mbox{ \boldmath $\xi$'},\mbox{ \boldmath $\eta$'})
\eeq
Where ${\bf P}_{\pi}$, ${\bf P}$ and ${\bf P}'$ are the momenta of the pion and the two nucleons respectively and
\beq
\mbox{ \boldmath $\xi$}=\frac{1}{\sqrt{2}}({\bf p}_1-{\bf p}_2)~,~
\mbox{ \boldmath $\eta$}=\frac{1}{\sqrt{6}}({\bf p}_1+{\bf p}_2-2{\bf p}_3)~,~
Q=\frac{1}{\sqrt{3}}({\bf p}_1+{\bf p}_2+{\bf p}_3)
\eeq
\beq
\mbox{ \boldmath $\xi$'}=\frac{1}{\sqrt{2}}({\bf p}'_1-{\bf p}'_2)~,~
\mbox{ \boldmath $\eta$'}=\frac{1}{\sqrt{6}}({\bf p}'_1+{\bf p}'_2-2{\bf p}'_4)~,~
Q'=\frac{1}{\sqrt{3}}({\bf p}'_1+{\bf p}'_2+{\bf p}'_4)
\eeq
\beq
\mbox{ \boldmath $\rho$}=\frac{1}{\sqrt{2}}({\bf p}'_3-{\bf p}_4)~,~Q_{\pi}=\frac{1}{\sqrt{2}}({\bf p}'_3+{\bf p}_4)
\eeq
Where $\bf{p}_i,~i=1,3$ are the momenta of the three quarks of one nucleon, ${\bf p}'_1,{\bf p}'_2,{\bf p}'_4$ the momenta of the three quarks of the other nucleon and ${\bf p}_4,{\bf p}'_3$ are those of the quarks involved in the pion. This notation was chosen since the interaction preserves the fermion lines ${\bf p}_i\longleftrightarrow  {\bf p}'_i$

 The above wave functions were normalized in the usual way:
 \beq
 \prec \psi_{{\bf P}}|\psi_{{\bf P}'}\succ =(2 \pi) ^3 \delta({\bf P}-{\bf P}')
~,~\prec \psi_{{\bf P}_{\pi}}|\psi_{{\bf P}'_{\pi}}\succ = 2E_{\pi} (2 \pi)^3 \delta({\bf P}_{\pi}-{\bf P}'_{\pi})
\label{wf}
\eeq
The internal wave functions are given by:
\beq
\phi(\xi)=\phi(0) e^{-(b^2_N\xi^2)/2},~~\phi(0)=\sqrt{\frac{b^3_N}{\pi \sqrt{\pi}}} \mbox { etc }
\eeq
The pion wave function has already been defined (see Eq. (\ref{Eq:pionwf})), except that
sometimes we will write:
\beq
\phi_{\pi}(0)=\phi(0) x^{3/2},~~x=\frac{b_{\pi}}{b_N}
\eeq
The integrals over the momentum variables ${\bf Q,Q^{'}}$ and ${\bf Q}_{\pi}$
 can be trivially performed due to the $\delta$ functions. Thus the orbital integral becomes:
\beq
 I_{\beta\beta}=(2 \pi )^3 \delta ({\bf P}-{\bf P}^{'}-{\bf P}_{\pi} ) {\cal M}
\eeq
\beq
 {\cal M} =\frac{(2 \pi )^{3/2}
\sqrt{2 E_{\pi}}}{3 \sqrt{3} (2 \sqrt{2})^{1/2}}
\int d^3 \mbox{ \boldmath $\xi$ }  d^3 \mbox { \boldmath $\xi$'}
 d^3 \mbox{ \boldmath $\eta$} d^3  \mbox{ \boldmath $\eta$'} d^3 \mbox{ \boldmath $\rho$}
\phi (\mbox{ \boldmath $\xi$},\mbox{ \boldmath $\eta$})
 \phi (\mbox { \boldmath $\xi$'},\mbox{ \boldmath $\eta$'})
 \phi_{\pi}( \mbox{ \boldmath $\rho$} ) \Omega_{\beta\beta}
 \eeq
 where $\Omega_{\beta\beta} $ depends on the mechanism involved as we now discuss.
\begin{enumerate}
\item The $q \bar{q}$ pair is created by the $0 \nu \beta \beta $ operator ( $0 \nu ~q \bar{q}$ case)\\
The case in which the $q \bar{q}$ pair is created by the $0 \nu \beta \beta $ operator (see fig. \ref{fig:piond}). Then up to terms linear in the momentum the effective operator takes the form:
 \beq
 \omega_{S(V)}=\sigma_4. \left (\frac{{\bf p}_4}{2 m_d}+\frac{{\bf p}^{'}_4}{2 m_u} \right ) \mbox{ (scalar and vector) }
\eeq
 \beq
 \omega_P=\sigma_3. \left (\frac{{\bf p}^{'}_3}{2 m_u} -\frac{{\bf p}_3}{2 m_d}\right ) \mbox{ (pseudoscalar ) }
\eeq
 \beq
 \omega_A=i(\sigma_3\times \sigma_4). \left (\frac{{\bf p}^{'}_4}{2 m_u}-\frac{{\bf p}_4}{2 m_d} \right ) \mbox{ (Axial current) }
\eeq
It is, of course, understood that the scalar and pseudoscalar must be multiplied by suitable coupling constants.
The full operator takes the form:
\beq \Omega_{\beta \beta \pi}=\frac{1}{(2 \pi )^3} \delta \left (p_3-p_3^{'}-p_4-p_4^{'}\right )
\delta\left( p_1-p_1^{'}\right )
~\delta \left ({\bf p}_2-{\bf p}^{'}_2\right ) \omega_i,~~i,S,V,P,A
 \eeq
The product of the three  $\delta$ functions can be cast in the form
$$ \delta \left (P-P^{'}-P_{\pi} \right )\delta \left (p_1 -p_1^{'}\right ) \delta \left (p_2-p_2^{'}\right )=$$
$$\delta \left (P-P^{'}-P_{\pi} \right )\delta \left (\sqrt{2}(\xi -\xi^{'})\right )\delta \left (\frac{1}{\sqrt{6}}(\eta -\eta^{'})+\frac{{\bf q}}{3}\right )$$
By setting $\xi^{'}=\xi$ and $\eta^{'}=\eta+\sqrt{\frac{2}{3}} {\bf q}$ we get
 \beq
 \omega_{S(V)}=\sigma_4. \left (
\frac{4 \left(3 \left(q-\sqrt{2}
   \rho \right) m_d+m_u \left(-5 q-2 \sqrt{6}
   \eta +2 p_N\right)\right)}{6 m_d m_u}  \right )
\eeq
 \beq
 \omega_P=\sigma_3. \left (
-\frac{\sigma_3 \left(m_d \left(q-2
   \sqrt{6} \eta +2 p_N\right)-3
   \left(q+\sqrt{2} \rho \right) m_u\right)}{6
   m_d m_u} \right ) \mbox{ (pseudoscalar ) }
\eeq
 \beq
 \omega_A=i(\sigma_3\times \sigma_4). \left (
-\frac{\sigma_4 \left(3
   \left(q-\sqrt{2} \rho \right) m_d+m_u \left(5
   q+2 \sqrt{6} \eta -2 p_N\right)\right)}{6 m_d
   m_u}\right ) \mbox{ (Axial) }
\eeq
After the integration (see next section) we get:
 \beq
 \omega_{S(V)}=\sigma_4. \left (
 \frac{\left (3 {\bf q} m_d+m_u
   \left(2 {\bf p}_N-3 {\bf q}\right)\right )}{6 m_d m_u} \right) \Rightarrow \frac{\sigma_4.{\bf p}_N}{m_p}
    \label{Eq:omegaSV}
\eeq
 \beq
 \omega_P=\sigma_3. \left (\frac{\left(3 {\bf q} m_u-m_d
   \left(3 {\bf q}+2 {\bf p}_N\right)\right)}{6 m_d m_u} \right ) \Rightarrow -\frac{\sigma_3.{\bf p}_N}{m_p}
    \label{Eq:omegaP}
\eeq
 \beq
 \omega_A=i(\sigma_3\times \sigma_4). \left ( \frac{ \left(3 {\bf q}
   \left(m_d+m_u\right)-2 m_u {\bf p}_N\right)}{6 m_d
   m_u} \right ) \Rightarrow i(\sigma_3\times \sigma_4).\frac{-3{\bf q}+{\bf p}_N}{m_p}
   \label{Eq:omegaA}
\eeq
The last expressions result in the case of the constituent mass for the quarks, $m_u=m_d=m_p/3$.
In the above equations:
\beq
 {\bf p} _N=\frac{{\bf P}+{\bf P}'}{2}~,~{\bf q}={\bf P}-{\bf P}'={\bf P}_{\pi}
\eeq
 \item Double beta decay and strong $q{\bar q}$ production ($^3P_0 ~q \bar{q}$ case).\\
In this case one needs the
collaborative effect of the $0\nu\beta\beta$ interaction acting between quarks together the strong interaction, which creates a pion out of
the vacuum (a' la $3P_0$ model or multigluon exchange):
 \beq H=g_r '\mbox{ \boldmath
$\sigma$}_4.{\bf B} ~\delta({\bf p}_4+{\bf p}'_4)~,~{\bf B}={\bf
p}_4-{\bf p}'_4
 \eeq
where $g'_r$ a dimensionless constant
proportional to the parameter $g_r=13.4\pm 0.1$, which is known
from experiment.
 One finds
 \beq
g_r^{'}=g_r
\frac{3 \sqrt{3} \left(2
   x^2+3\right)^{3/2}}{80
   \sqrt[4]{2} \phi_{  \pi}(0) \pi^{3/2}
   m_p \sqrt{m_{\pi }}}
\eeq
 where
5where  $\phi_{\pi}(0)$ is the pion wave function at the origin.
 \begin{itemize}
\item The direct term in the one pion contribution.\\
In this case (see fig. \ref{fig:pionb})  none of the two interacting quarks participates in
the pion as defined above.  Thus we get:
\beq \Omega_{\beta \beta \pi}=\frac{g^{'}_r}{(2 \pi )^3} \delta \left (p_1+p_2-p_1^{'}-p_2^{'}\right )
\delta\left( p_3-p_4^{'}\right )
~\delta \left ({\bf p}_4+{\bf p}^{'}_4 \right ) \sigma_4.\left ({\bf p}_4+{\bf p}^{'}_4 \right )
 \eeq
The product of the above three  $\delta$ functions can be cast in the form
$$ \delta \left (P-P^{'}-P_{\pi} \right )\delta \left (p_1+p_2-p_1^{'}-p_2^{'}\right ) \delta \left ({\bf p}_4+{\bf p}^{'}_4 \right ) $$
The first of these $\delta$-functions  expresses momentum conservation. Going into the Jacobi variables we find:
 \barr
 \Omega_{\beta \beta \pi}&=&\frac{1}{(2 \pi )^3}
 \delta(P-p^{'}-P_{\pi})\delta \left (
\frac{2 q+\sqrt{6} (\eta
   -\eta^{'})}{3}  \right )
\nonumber\\
&&\delta \left (
\frac{2 q-2 \sqrt{6}
   \eta ^{'}-3 \sqrt{2} \rho +2
   p_N }{6}\right ) \sigma_4.
\frac{4 q+2 \sqrt{6}
   \eta^{'}-3 \sqrt{2} \rho -2
   p_N}{6}
   \earr

We find it convenient to use the above $\delta$ functions to obtain:
\beq
\eta=-\frac{2 q+3 \sqrt{2} \rho -2 p_N}{2
   \sqrt{6}},~~\eta^{'}=
\frac{2 q-3 \sqrt{2} \rho +2 p_N}{2
   \sqrt{6}}
   \eeq
    One finds:
 \beq
 \mbox{ \boldmath $\sigma$ }_4.{\bf B}=\mbox{ \boldmath $\sigma$ }_4. \left ({\bf q}-\sqrt{2}\mbox{ \boldmath $\rho$ } \right )
 \eeq
 Furthermore  A-terms, appearing in the case of the pseudoscalar contribution, take the form:
 \beq
 {\bf A}_1= -\frac{\left(m_u\left(\sqrt{2} (\rho - 2\xi) - 2 p_N\right) +
              m_d\left(\sqrt{2} (2 \xi^{'} - \rho) +                  2 p_N\right)\right)\sigma _1}{4 m_d m_u}
 \eeq
 \beq
  {\bf A}_2= = -\frac{\left(m_u\left(\sqrt{2} (2\xi + \rho) - 2 p_N\right) +
              m_d\left(2 p_N - \sqrt{2} (2\xi^{'} + \rho)\right)\right)\sigma \
_2}{4 m_d m_u}
 \eeq
  Thus using the corresponding $\delta$-functions the $\eta$ and $\eta^{'}$ integrations can be done trivially.
 \item The exchange term in the one pion contribution.\\
 By this we mean that one of the interacting particles participates in the pion (see fig. \ref{fig:pionc}) .
 Proceeding as above have:
 \beq \Omega_{\beta \beta \pi}=\frac{g^{'}_r}{(2 \pi )^3} \delta \left (p_2+p_3-p_2^{'}-p_3^{'}\right )
\delta\left( p_1-p_1^{'}\right )
~\delta \left ({\bf p}_4+{\bf p}^{'}_4 \right ) \sigma_4.\left ({\bf p}_4+{\bf p}^{'}_4 \right )
 \eeq
 Going into the Jacobi variables we find:
 \barr
 \Omega_{\beta \beta \pi}&=&\frac{1}{(2 \pi )^3}
 \delta(P-P^{'}-P_{\pi})\delta \left (
\frac{2 p_N-\sqrt{2} \left(\sqrt{3}
   \eta +\sqrt{3} \eta^{'}+3 (\xi
   -\xi^{'}+\rho )\right )}{6}  \right )
\nonumber\\
&&\delta \left (
\frac{2 q-2 \sqrt{6} \eta^{'}
   -3 \sqrt{2} \rho +2 p_N}{6} \right )
 \sigma_4.({\bf q}-\sqrt{2} \rho)
   \earr

We find it convenient to use the above $\delta$ functions to obtain:
$$
\xi^{'}=
\frac{1}{6} \left(\sqrt{2} q+2 \sqrt{3} \eta +6
   \xi +3 \rho -\sqrt{2} p_N\right),~~\eta^{'}=
\frac{2 q-3 \sqrt{2} \rho +2 p_N}{2 \sqrt{6}}$$
 Thus the $\xi^{'}$ and $\eta^{'}$ can be done trivially. Furthermore  A-terms, appearing in the case of the pseudoscalar contribution, take the form:
 \beq
  {\bf A}_2=
\frac{\sigma_2 \left(m_d
   \left(q+\sqrt{2} \left(\sqrt{3} \eta +3 (\xi
   +\rho )\right)-4 p_N\right)+m_u
   \left(q+\sqrt{6} \eta -3 \sqrt{2} \xi +2
   p_N\right)\right)}{6 m_d m_u}
 \eeq
  \beq
 {\bf A}_3=
\frac{\sigma_3 \left(m_u \left(q-2
   \sqrt{6} \eta +2 p_N\right)-3
   \left(q+\sqrt{2} \rho \right) m_d\right)}{6
   m_d m_u}
 \eeq
 \end{itemize}
\end{enumerate}
\begin{table}[t]
\caption{ The spin flavor matrix elements of the various spin operators encountered in this work. They are normalized to the matrix element of the nucleon spin.
}
\label{table.spinme}
\begin{center}
\begin{tabular}{|c|c|c|}
\hline
 &   &   \\
$\Omega_{s}$ &process& $ME{sf}=\frac{<|\Omega_{s}|>}{<|\sigma_N|>}$\\
( k indicates the spin ranks) & & \\

 \hline
$\sigma_4$&scalar or vector&-$\frac{5 \sqrt{2}}{9}$\\
 \hline
 $\sigma_3$ &pseudoscalar&-$\frac{5 \sqrt{2}}{9}$\\
 \hline
 $i\sigma3\times \sigma_4$&axial&$\frac{10 \sqrt{2}}{9}$\\
 \hline
 $\sigma_4$&direct&$-\frac{ \sqrt{2}}{9}$\\
 \hline
 $(\sigma_1.\sigma_2)\sigma_4$&direct&$-\frac{ \sqrt{2}}{9}$\\
 \hline
 $\left[(\sigma_1 \times \sigma_2)k_{12}=2;\sigma_4 \right]k=1$&direct&$\frac{4\sqrt{10}}{9 \sqrt{3}}$\\
  \hline
 $(\sigma_1.\sigma_4)\sigma_2$&direct&$-\frac{ 7\sqrt{2}}{9}$\\
 \hline
  $(\sigma_2.\sigma_4)\sigma_1$&direct&$-\frac{ 7\sqrt{2}}{9}$\\
  \hline
  $\sigma_4$&exchange&$\frac{ \sqrt{2}}{9}$\\
 \hline
  $(\sigma_2.\sigma_3)\sigma_4$&exchange&$\frac{ \sqrt{2}}{9}$\\
 \hline
 $\left[(\sigma_2 \times \sigma_3)k_{23}=2;\sigma_4 \right]k=1$&exchange&$\frac{8\sqrt{10}}{9 \sqrt{3}}$\\
  \hline
  $(\sigma_2.\sigma_4)\sigma_3$&exchange&$-\frac{13 \sqrt{2}}{9}$\\
   \hline
  $(\sigma_3.\sigma_4)\sigma_2$&exchange&$-\frac{ 13\sqrt{2}}{3}$\\
 \hline
\end{tabular}
\end{center}
\end{table}
 \subsection{The $0\nu\beta\beta$ decay amplitude at the nucleon level.}
Performing the orbital integrals we encountered in the previous section, we must evaluate the spin-flavor ME for the various operators encountered above, classified according to their spin rank. The obtained matrix elements, in units of the nucleon spin ME are included in  table \ref{table.spinme}). Using these results  one can obtain the needed amplitude at the nucleon level.
 As  expected from the above discussion we will consider three possibilities:
\begin{enumerate}
\item The $0\nu ~q \bar{q}$ case\\
In this case we can write the amplitude as
\beq
 {\cal M} =  \frac{1}{(2 \pi)^{3/2}} \frac{1}{3 \sqrt{3}}\frac{\sqrt{2 m_{\pi}}}{\sqrt{2 \sqrt{2}}}\mbox{ \boldmath $\sigma$}_N. {\bf C}_i~ ME(sf) J_{orb}
 \eeq
where $\mbox{ \boldmath $\sigma$}_N$  is the nucleon spin and $ME(s-f)$ is the spin-flavor matrix element
( see table \ref{table.spinme}) and $J_{orb}$ is the radial integral. One finds
\beq
J_{orb}=6 \sqrt{6}\frac{\phi_{\pi}(0)}{(\phi(0))^2}e^{-(b^2_Nq^2)/6}
\eeq
The coefficients ${\bf C}_i$ can be read off from Eqs \ref{Eq:omegaSV}-\ref{Eq:omegaA}, namely
\beq{\bf C}_{S(V)}= \left (
 \frac{\left (3 {\bf q} m_d+m_u
   \left(2 {\bf p}_N-3 {\bf q}\right)\right )}{6 m_d m_u} \right) \Rightarrow \frac{{\bf p}_N}{m_N}
\label{Eq:CSV}
\eeq
\beq  {\bf C}_{P}= \left (\frac{\left(3 {\bf q} m_u-m_d
   \left(3 {\bf q}+2 {\bf p}_N\right)\right)}{6 m_d m_u} \right ) \Rightarrow -\frac{{\bf p}_N}{m_N}
\label{Eq:CP}
\eeq
 \beq
 {\bf C}_{A}= \left ( \frac{ \left(3 {\bf q}
   \left(m_d+m_u\right)-2 m_u {\bf p}_N\right)}{6 m_d
   m_u} \right ) \Rightarrow \frac{3{\bf q}-{\bf p}_N}{m_N}
\label{Eq:CA}
\eeq
The term $p_N$ of the amplitude will lead to a non local effective operator in coordinate space.
 \item The $^3P_0 ~q \bar{q}$ case\\
 Double beta decay proceeds via two quarks in a state with isospin one, which is  color antisymmetric. So the two quarks must be  in a spin one state. So there is no contribution in V-A theories, since
the vector and the axial vector contributions are identical. For the scalar and pseudoscalar
cases the needed couplings depend on the particle model assumed. In the R-parity violating SUSY the coupling is , e.g.
$\frac{3}{8}(\eta^T+\frac{5}{3}\eta_{PS})$ found in \cite{FKS98}. In our discussion we will not include such a
model dependent coupling.
 We will
distinguish the two possibilities:

  {\bf a) The direct term.}\\
In this case we can write the amplitude as
\beq
 {\cal M} =  \frac{1}{(2 \pi)^{3/2}} \frac{1}{3 \sqrt{3}}\frac{\sqrt{2 m_{\pi}}}{\sqrt{2 \sqrt{2}}}
{\bf A}_1.{\bf A}_2 J_{orb}
 \eeq
 In the case of the scalar contribution we find
from table \ref{table.spinme} that
\beq
{\bf A}_1.{\bf A}_2=-\frac{\sqrt{2}}{9}{\bf q}.\sigma_N
\eeq
In the case of the pseudoscalar contribution (see Appendix) and in the local approximation ${\bf p}_N=0$ we find:
\beq
{\bf A}_1.{\bf A}_2=\frac{1}{3} \left (\frac{\left(m_d-m_u \right)}
{\sqrt{4 x^2+6} b_N m_d m_u} \right )^3 (\sigma_1.\sigma_2)q.\sigma _4
\eeq
We expect this to be a good approximation. In any event it makes the operator tractable.

The corresponding orbital integral is:
\barr
J_{orb}&=&g_r^{'}\frac{2^3 3^2}{(3+2 x^2)\sqrt{3+2 x^2}}\frac{\phi_{\pi}(0)}{(\phi(0))^2}e^{-b_N ^2 q^2/6}
e^{-b^2 _N p_N^2 \left ((2 x^2)/(3+2 x^2) \right)/6 }=
\nonumber\\
&& g_r
\frac{81 \sqrt{3}}
   {10
   \sqrt[4]{2} \phi^2(0)\pi^{3/2}}
e^{-b^2 _N p_N^2 \left ((2 x^2)/(3+2 x^2) \right)/6 }
\label{Eq:gr}
\earr
We not with satisfaction that any uncertainties in the pion w.f. have dropped out, at least if the non local
term in the exponential are ignored.

{\bf b) The exchange term.}\\
The amplitude takes the form:
\\\beq
 {\cal M} =  \frac{1}{(2 \pi)^{3/2}} \frac{1}{3 \sqrt{3}}\frac{\sqrt{2 m_{\pi}}}
 {\sqrt{2 \sqrt{2}}}{\bf A}_2.{\bf A}_3 J_{orb}
 \eeq
 Again there is no contribution in V-A theories, since
the vector and the axial vector contributions are identical. In the case of the scalar contribution we find
from table \ref{table.spinme} that
\beq
{\bf A}_2.{\bf A}_3=\frac{\sqrt{2}}{9}{\bf q}.\sigma_N
\eeq
In the case of the pseudoscalar contribution  for the constituent quark masses we get:
    \beq
{\bf A}_2.{\bf A}_3=
\left[q^2\frac{320 \sqrt{2}  \left(7 x^2+1\right) \left(56
   x^2+3\right)^2}{147 \left(28 x^2+3\right)^3 m_N^2}
+
\frac{416 \sqrt{2} \left(588 x^4-77 x^2+57\right)}{63
   \left(28 x^2+3\right)^2 b_N^2 m_N^2}\right]
 \sigma_N.q
   \eeq
where $x=\frac{b_{\pi}}{b_N}$.
  Note the presence of the $q^2$ in the first term. This will lead to an operator with a different
radial dependence, i.e.  $F^{(k)}_i(x)\Longrightarrow -\nabla ^2  F^{(k)}_i(x)$ (see Eq. (\ref{Eq:Frs})).
   The corresponding orbital integral for the exchange term is:
\beq
J_{orb}=g_r^{'}\frac{ 3^3~2^7 \sqrt{2} }{(3+28 x^2)\sqrt{3+28 x^2}}\frac{\phi_{\pi}(0)}{(\phi(0))^2}
e^{-b_N ^2 \left(({\bf q}-{\bf p}_N)^2/6\right)\left((4x^2)/(3+28x^2) \right)}
\eeq
or
\beq
J_{orb}=g_r
\frac{648 \sqrt[4]{2} \sqrt{3}
   \left(2 x^2+3\right)^{3/2}
   }{5 \pi ^{3/2} \left(28
   x^2+3\right)^{3/2} \phi^2(0) m_p \sqrt{m_\pi}}
e^{-b_N ^2 \left(({\bf q}-{\bf p}_N)^2/6\right)\left((4x^2)/(3+28x^2) \right)}
\eeq
In this instance the obtained results depend on the pion w.f. at the origin (via x).
\end{enumerate}
\section{Results}
Our main results are the coefficients $\alpha_{2\pi}$ and
$\alpha_{1\pi}$, which multiply the standard nuclear matrix elements. We will not  elaborate further on the new non local terms (at the nucleon level). 
\subsection{The coupling coefficients $\alpha_{2\pi}$}
 Before presenting our results we should mention that in the elementary particle treatment \cite{FKS98} one
  can write
  \beq
  \alpha_{2 \pi}=\frac{1}{6 f^2_A}g^2_rh^2_{\pi}\left( \frac{m_{\pi}}{m_p}\right)^4
  \eeq
  Obtained under the factorization approximation:
  \beq
 < \pi^+|J_Pj_P|\pi^- >=\frac{5}{3}< \pi^+|J_P|0><0|j_P|\pi^- >,~~<0|J_P|\pi^->=m^2_{\pi}h_{\pi}
 \label{Eq:EP2pi}
  \eeq
  The parameter $h_{\pi}$ is given by
  \beq
  h_{\pi}=i \sqrt{2} 0.668 \frac{m_{\pi}}{m_d+m_u}
  \eeq
  Returning back to our approach we note that the non relativistic reduction is applicable in the constituent quark mass framework, $m_u=m_d \approx m_N/3$. In this case the pseudoscalar term contribution becomes:
$$
\alpha_{2 \pi}=-0.0005 \mbox{ (for x=1.0) },~~\alpha_{2 \pi}=-0.05\mbox{ (for x=0.4)}$$
We should compare this with the value obtained in V-A theory, see Eq. (\ref{alpha2pi}), using $f^2_{\pi N N}=0.08 $ and $b_N=1.0$ fm:
$$\alpha_{2 \pi}=0.013~~ ( x=1) \mbox{ and } \alpha_{2 \pi}=0.11~~ (x=1.0)$$
i.e. it  is quite a bit smaller. It is also
much smaller than the  value
0.20 obtained in the elementary particle treatment \cite{FKS98} using current quark masses. This disagreement
cannot be healed by the fact that in the present
case we encounter a very strong dependence of the results on the pion size parameter, see  Fig. \ref{fig:2piP},
unless we use very unrealistic values of the pion size parameter.
 One expects, of course, an enhancement of the pseudoscalar contribution, if one uses the current
  quark masses, since they  are assumed to be very small.  Indeed this way for typical values $x=1$,
 $b_N=1$ fm, $m_d=5$ MeV and
  $m_u=10$ MeV we obtain $\alpha_{2 \pi}=-1.3$ and $\alpha_{2 \pi}(\Omega_{1 \pi})=0.08$,
 which are  very large. We should mention, however, that the validity of  the non
 relativistic reduction at the quark level may be questionable in
this case.
\subsection{The coupling coefficients $\alpha_{1\pi}$}
Before proceeding further we will briefly present how the coefficient $\alpha_{1\pi}$
 was obtained in the context
of the elementary particle treatment \cite{FKS98}:
\beq
\alpha_{1\pi}=-F_P \frac{1}{36 f^2_A}g_rh_{\pi}\left( \frac{m_{\pi}}{m_p}\right)^4
  \eeq
 The needed parameters were obtained using the factorization approximation one writes in  the case of $1-{\pi}$ mode
\beq
<p|j_P J_P|n\pi_>=\frac{5}{3}<p|J_P|n><0|J_P|\pi^->,~~<p|J_P|n>=F_P\approx4.41
 \label{Eq:EP1pi}
\eeq
The matrix element $<0|J_P|\pi^->$ was given above (see Eq. (\ref{Eq:EP2pi})). Thus these authors \cite{FKS98}
 find:
\beq
\alpha_{1\pi}=-4.4\times 10^{-2}
\label{Eq:alpha1pi}
\eeq

\begin{figure}[!ht]
 \begin{center}
\includegraphics[scale=0.6]{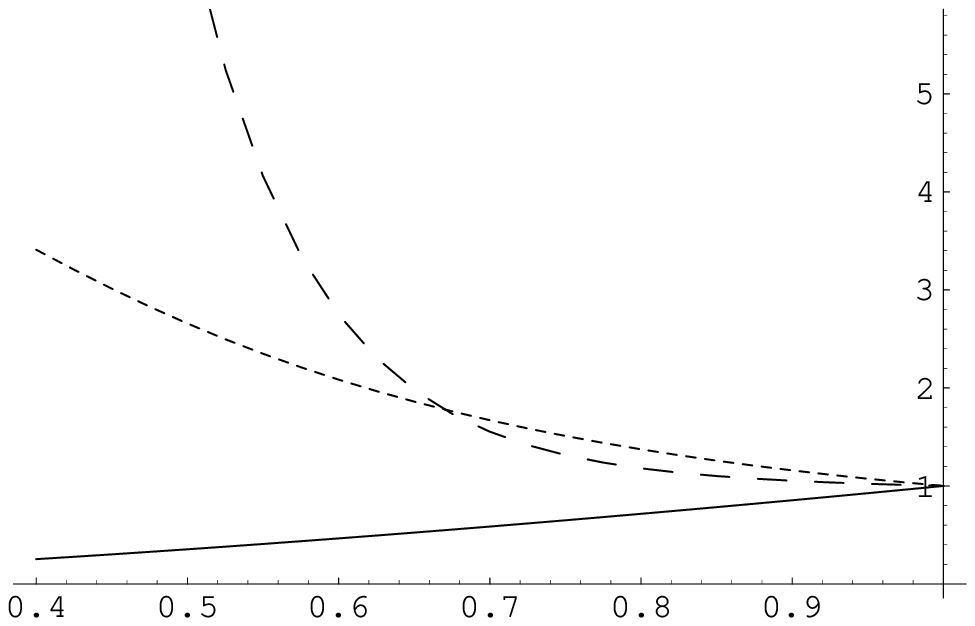}
\includegraphics[scale=0.6]{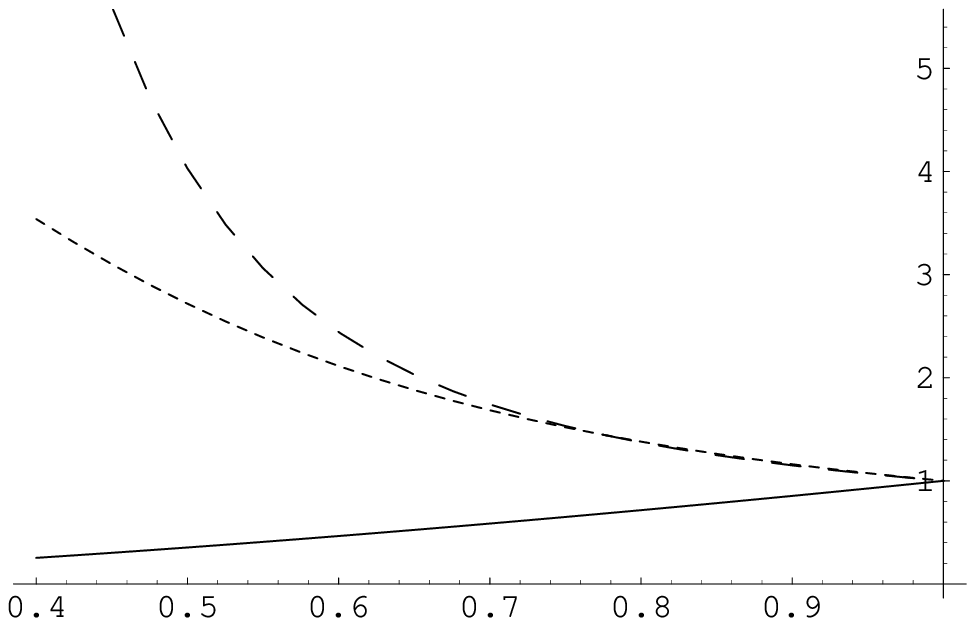}\\
\hspace{-0.0cm} {$x \longrightarrow$}
 \caption{ The functions which provide the dependence of $1-{\pi}$ amplitude on the pion size parameter through the variable $x=b_{\pi}/b_N$ are exhibited. On the left we show the relevant coefficients using the current quark masses. The continuous curve is associated with the
coefficient $f^A_{1\pi}$ ( see Eq. (\ref{Eq:fA})), the long dash is associated with with the exchange q-independent coefficient ($f^{cur}_{1\pi}$) and the the short dash with that of $g^{cur}_{1\pi}$
(see Eq. (\ref{Eq:cur}) ). On the right we show the same quantities obtained with constituent quark masses.}
    \label{fig:current}
   \end{center}
  \end{figure}

Returning to our approach  these coefficients are obtained in the following procedure: First we write
\beq
\frac{4 \pi}{m_p^2} {\cal M}=c_{1 \pi} g_r \frac{\sigma_N.{\bf q}}{2 m_p}
\eeq
Then, ignoring the momentum dependence in the exponential, we get:
\begin{enumerate}
\item Double beta decay only.\\
From Eqs(\ref{Eq:CSV})-(\ref{Eq:CA}) we see that the only local contribution comes from the axial current.
\beq
c_{1\pi}=\frac{10 \sqrt{2} \sqrt[4]{\pi
   }  \sqrt{m_{\pi }}
   \left(m_d+m_u\right)}{9
   g_r \sqrt{b_N^3} m_d
   m_N m_u}f^A_{1\pi}(x) \mbox{ (current masses)}
\eeq
\beq
c_{1\pi}=\frac{20 \sqrt{2} \sqrt[4]{\pi
   }  \sqrt{m_{\pi }}}{3
   g_r \sqrt{b_N^3}
   m_N^2} f^A_{1\pi}(x) \mbox{ (constituent masses)}
\eeq
with
\beq
f^{A}_{1\pi}(x)=x^{3/2}
\eeq
Using $m_d=5$ MeV, $m_u=10$ MeV and $g_r=13.4$ we get
\beq
c_{1\pi}=1.6~f^{A}_{1\pi}(x), \mbox{ (current masses )}
\eeq
 On the other hand for the constituent masses we find:
\beq
c_{1\pi}=3.4\times 10^{-2}~~f^{A}_{1\pi}(x), \mbox{ (constituent masses )},
\eeq

The corresponding coefficient that must multiply the nuclear matrix element is $\alpha_{1\pi}$
\beq
\alpha_{1\pi}=c_{1\pi} \frac{f^2_{\pi NN}}{f^2_A}
\eeq
\beq
\alpha_{1\pi}=0.085~f^{A}_{1\pi}(x) \mbox{ (current masses )},\alpha_{1\pi}=1.8\times10^{-3}~f^{A}_{1\pi}(x) \mbox{ (constituent masses )}
\label{Eq:fA}
\eeq
\item The direct term\\
There is no contribution of the direct diagram if the non local terms are ignored.
\item the exchange term\\
\begin{itemize}
\item In the case of the current quark masses we get the standard term:\\
\beq
c_{1\pi}=1.0\times 10^3 f^{cur}_{1\pi}(x)
\eeq
In addition we have an operator which results from the term in the amplitude, which was cubic in $q$. Thus
we factor out the $q^2 m^2_{\pi}$ and absorb it in the effective transition operator. In the remaining coefficient  we merely replace  $q^2$ by $m^2_{\pi}$. Thus
\beq
c_{1\pi}=91.5 g^{cur}_{1\pi}(x)
\eeq
Proceeding as above we get respectively:
\beq
\alpha_{1\pi}=51~f^{cur}_{1\pi}(x)~~\mbox{or}~~\alpha_{1\pi}=51~g^{cur}_{1\pi}(x)
\label{Eq:cur}
\eeq
The coefficient $f^{cur}_{1\pi}(x)$ is associated with the standard operator $\Omega_{1\pi}(x_{\pi})$, while
$g^{cur}_{1\pi}(x)$ must be linked with a new type of operator $\tilde{\Omega}_{1\pi}(x_{\pi})$ with modified radial dependence , i.e.  $F^{(k)}_i(x)\Longrightarrow -\nabla ^2  F^{(k)}_i(x)$ (see Eq. (\ref{Eq:Frs})). Both coefficients are so normalized that  $f^{cur}_{1\pi}(1)$=$g^{cur}_{1\pi}(1)=1$.
 In any case the use of current quark masses leads to very large values.\\
\item The constituent quark masses.\\
In this case  we get:
\beq
c_{1\pi}=1.37~ f^{con}_{1\pi}(x)\mbox{ or  }c_{1\pi}=1.72 g^{con}_{1\pi}(x)
\eeq
\beq
\alpha_{1\pi}=0.071~f^{con}_{1\pi}(x)\mbox{ or  }\alpha_{1\pi}=0.090~g^{con}_{1\pi}(x)
\label{Eq:con}
\eeq
Again the coefficient $f^{con}_{1\pi}(x)$ is associated with the standard operator%
, while
$g^{con}_{1\pi}(x)$ must be linked with the operator $\tilde{\Omega}_{1\pi}(x_{\pi})$, with $f^{con}_{1\pi}(1)=g^{con}_{1\pi}(1)=1$.
\end{itemize}
The functions $ f^{A}_{1\pi},f^{cur}{1\pi}(x)$,  $g^{cur}_{1\pi}(x)$  $ f^{A}_{1\pi},f^{con}_{1\pi}(x)$ and $g^{con}_{1\pi}(x)$are shown in Fig. \ref{fig:current}.
 For $x=1$ for the standard local  $1\pi$ operator considering all contributions mentioned above with constituent quark masses we find 
$\alpha_{1\pi}=7.3\times10^{-2}$, which is in size almost a factor of 2 larger  than that obtained in elementary particle treatment \cite{FKS98} (see Eq.(\ref{Eq:alpha1pi})) . Note, however, that our results  depend on the pion size parameter.
\end{enumerate}
\section{discussion}
 In the present paper we have considered the effective $0\nu \beta \beta$ decay operator associated with the exchange of heavy particles mediated by pions in flight between nucleons. A harmonic oscillator non relativistic quark model in momentum space  was employed for the pion and the nucleon. This allowed one to separate out the relative from the center of mass motion. The ratio of the pion to the nucleon harmonic oscillator parameter, $x=b_{\pi}/b_{N}$ was treated as a parameter. When needed, the constituent quark mass equal to 1/3 of the nucleon mass employed.\\
The obtained results were compared to the elementary particle treatment, with current quark masses, previously employed.\\
In the case of the two pion mode  we find a new term with different momentum dependence, which is not present in the elementary particle treatment. This gives rise to a new operator, which has the same structure as the one previously associated with the one pion mechanism.

In connection with one  pion mechanism we found that there exist three diagrams, which cannot be distinguished in the elementary particle treatment, namely:
\begin{enumerate}
\item Diagrams in which the $q{\bar q}$ is crated out of the vacuum via the strong interaction.\\
 In this case we employed the  $^3P_0$ model.
The strength of this interaction was fitted to the pion nucleon coupling $g_r$.\\
We distinguished two possibilities:
\begin{itemize}
\item The two interacting quarks participate only in the structure of the  nucleon.
\item One of the interacting quarks participates in the structure of the pion.
\end{itemize}
\item The the $q{\bar q}$ is crated by the weak interaction itself.
\end{enumerate}
 Depending on the mechanism we encountered new non local terms, i.e. terms which depend on the nucleon momentum. These will lead to new types of effective nuclear operators, which have not been  examined up to now.

   The results obtained in the present calculation depend among other things on the ratio of the pion
   to nucleon size parameters. Using reasonable values for this ratio we obtain values of $\alpha_{1\pi}$, which are in good agreement with those obtained in the
elementary particle treatment. Regarding the couplings  $\alpha_{2\pi}$, however, we find that they 
 are slightly smaller than those obtained in the elementary particle
treatment in the case of the V-A theory. They are, however, quite a bit smaller than those 
obtained in the case of the pseudoscalar term, when the
constituent quark masses are used.  We can,
of course, obtain much larger values for the pseudoscalar term, if
the current quark masses are used.Admittedly, however, it may not be very
consistent to do so in our approach, since it is essentially a non
relativistic treatment. We thus suspect that the small current quark masses are behind the large values found in the elementary particle treatment. 

In summary,  taking into account the fact that a number of approximations are behind both approaches, we may say that there exists a reasonable agreement between them, which gives a degree of  confidence in both. A more complete comparison can, of course,  be made only after the inclusion in the calculation of the nuclear matrix elements of the new operators found in the present approach, namely: i) the local operator $\tilde{\Omega}_{1\pi}(x_{\pi})$ resulting from terms cubic in $q$ and ii) the non local operators, which depend on the nucleon momentum.
\section{Acknowledgments}
The work of one of us (JDV) started while he was visiting RCNP, it was continued while he was in Tuebingen under a Alexander von Humboldt Research Award and it was finished during a visit at CERN. He is indebted to  these institutions as well as to the grant MRTN-CT-2006-035863 (UniverseNet) for their support and to Professors H. Toki, A. Faessler and I. Antoniadis for their hospitality. He is also  happy to acknowledge useful discussions with H. Ejiri.
  \section{Appendix}
  In this appendix we will present the relevant formulas in the case the $q{\bar q}$ pair  is produced in
  via the strong interaction.
  In the case of the 1-pion contribution we get:
  \begin{enumerate}
  \item  The direct term\\
In this case we can write the amplitude as
\beq
 {\cal M} =  \frac{1}{(2 \pi)^{3/2}} \frac{1}{3 \sqrt{3}}\frac{\sqrt{2 m_{\pi}}}{\sqrt{2 \sqrt{2}}}
{\bf A}_1.{\bf A}_2 J_{orb}
 \eeq
 In the case of the scalar contribution we find
from table \ref{table.spinme} that
\beq
{\bf A}_1.{\bf A}_2=-\frac{\sqrt{2}}{9}{\bf q}.\sigma_N
\eeq
In the case of the pseudoscalar contribution we find:
\barr
{\bf A}_1.{\bf A}_2&=&
\frac{\left(x^2+2\right)
   \left(m_d-m_u\right) p_N .\sigma
   _1}{\left(2 x^2+3\right) m_d
   m_u}~\frac{\left(x^2+2\right)
   \left(m_d-m_u\right) p_N .\sigma
   _2}{\left(2 x^2+3\right) m_d
   m_u} \left(q+\frac{2 p_N}{2
   x^2+3}\right) .\sigma _4
   \nonumber\\
&&
\frac{1}{3}\frac{\left(x^2+2\right)
   \left(m_d-m_u\right) p_N \sigma
   _1}{\left(2 x^2+3\right) m_d
   m_u}
\frac{\left(m_d-m_u\right) }{\sqrt{2} \sqrt{2 x^2+3} b_N
   m_d m_u}
\frac{2 \sqrt{2} }{\sqrt{2 x^2+3} b_N}\sigma_2.\sigma_4
\nonumber\\
   &&
\frac{1}{3}\frac{\left(x^2+2\right)
   \left(m_d-m_u\right) p_N \sigma
   _2}{\left(2 x^2+3\right) m_d
   m_u}
\frac{\left(m_d-m_u\right) }{\sqrt{2} \sqrt{2 x^2+3} b_N
   m_d m_u}
\frac{2 \sqrt{2}}{\sqrt{2 x^2+3} b_N}\sigma_1.\sigma_4
\nonumber\\
&&+\frac{1}{3}\left(q+\frac{2 p_N}{2
   x^2+3}\right). \sigma _4
\left(\frac{\left(m_d-m_u\right)}{\sqrt{2} \sqrt{2 x^2+3} b_N
   m_d m_u} \right )^2 \sigma_1.\sigma_2
\earr
In the limit $m_d =m_u$ the above expression vanishes. In the local approximation ${\bf p}_N=0$ we find
\beq
{\bf A}_1.{\bf A}_2=\frac{1}{3} \left (\frac{\left(m_d-m_u \right)}
{\sqrt{4 x^2+6} b_N m_d m_u} \right )^3 (\sigma_1.\sigma_2)q.\sigma _4
\eeq
We expect this to be a good approximation. In any event it makes the operator tractable.
The corresponding orbital integral is
given by Eq. (\ref{Eq:gr}).
 \item The exchange term\\
The amplitude takes the form:
\\\beq
 {\cal M} =  \frac{1}{(2 \pi)^{3/2}} \frac{1}{3 \sqrt{3}}\frac{\sqrt{2 m_{\pi}}}{\sqrt{2 \sqrt{2}}}{\bf A}_2.{\bf A}_3 J_{orb}
 \eeq
 Again there is no contribution in V-A theories, since
the vector and the axial vector contributions are identical. In the case of the scalar contribution we find
from \ref{table.spinme} that
\beq
{\bf A}_2.{\bf A}_3=\frac{\sqrt{2}}{9}{\bf q}.\sigma_N
\eeq
In the case of the pseudoscalar contribution we find:
\barr
{\bf A}_2.{\bf A}_3&=&
-\frac{\left(7 x^2+1\right) \left(21 \
\left(14 x^2+1\right) m_d-\left(70 x^2+9\right) m_u\right) \left(\left(224 \
x^2-9\right) m_d+3 \left(224 x^2+19\right) m_u\right)}{441 \left(28 x^2+3
\right)^3 m_d^2 m_u^2}
\nonumber\\
&&q.\sigma _2 q.\sigma _3 q.\sigma _4
\nonumber\\
&&+
\frac{ \left(\left(1400 x^2+117
\right) m_d-3 \left(168 x^2+23\right) m_u\right) \left(\left(448 x^2+45 \right)
m_u-21 m_d\right)} {3528 \left (28 x^2+3 \right )^3 m_d^2 m_u^2}p_N.\sigma _2 p_N.\sigma _3 p_N.\sigma _4
\nonumber\\
&&+
\frac{4  \left(11 m_d+5 m_u\right) \left(21 \left(14 x^2+1\right)
m_d-\left(70 x^2+9\right) m_u\right) }{63 \left(28 x^2+3
\right)^2 b_N^2 m_d^2 m_u^2}\sigma _2 .\sigma _4~q.\sigma _3
\nonumber\\
&&+
\frac{2  \left(11 m_d+5 m_u\right) \left(21 m_d-\left(448 x^2+45\
\right) m_u\right) }{63 \left(28 x^2+3\right)^2 b_N^2
m_d^2 m_u^2}\sigma _2 .\sigma _4~p_N.\sigma _3
\nonumber\\
&&+
\frac{2 q.\sigma _2 \left(7 m_d+
            m_u\right) \left(\left(224 x^2-9\right) m_d+3 \left(224 x^2+19
\right) m_u\right) }{63 \left(28 x^2+3\right)^2 b_N^2 m_d^2
m_u^2}\sigma _3. \sigma _4~q.\sigma _2
\nonumber\\
&&
-\frac{2  \left(7 m_d+
            m_u\right) \left(\left(1400 x^2+117\right) m_d-3 \left(168 x^2+23
\right) m_u\right) }{63 \left(28 x^2+3\right)^2 b_N^2
m_d^2 m_u^2}\sigma _3 .\sigma _4~p_N.\sigma _2
\nonumber\\
&&
-\frac{8 \left(7 x^2+1\right)  \left(11
   m_d^2+\left(4 x^2+7\right) m_u m_d+2 \left(6
   x^2+1\right) m_u^2\right) \sigma _2 \sigma _3}{3
   \left(28 x^2+3\right)^2 b_N^2 m_d^2 m_u^2} \sigma _2. \sigma _3~q.\sigma _4
\nonumber\\
&&+
\frac{2 p_N.\sigma _4 \left(11 m_d^2+\left(4 x^2+7\right)
   m_u m_d+2 \left(6 x^2+1\right) m_u^2\right) }{3 \left(28 x^2+3\right)^2 b_N^2 m_d^2
   m_u^2}\sigma _2. \sigma _3~p_N.\sigma _4
\earr
In the limit of ignoring the non local terms we get:
\barr
{\bf A}_2.{\bf A}_3&=&
-\frac{\left(7 x^2+1\right) \left(21 \
\left(14 x^2+1\right) m_d-\left(70 x^2+9\right) m_u\right) \left(\left(224 \
x^2-9\right) m_d+3 \left(224 x^2+19\right) m_u\right)}{441 \left(28 x^2+3
\right)^3 m_d^2 m_u^2}
\nonumber\\
&&q.\sigma _2 q.\sigma _3 q.\sigma _4
\nonumber\\
&&+
\frac{4  \left(11 m_d+5 m_u\right) \left(21 \left(14 x^2+1\right)
m_d-\left(70 x^2+9\right) m_u\right) }{63 \left(28 x^2+3
\right)^2 b_N^2 m_d^2 m_u^2}\sigma _2 .\sigma _4~q.\sigma _3
\nonumber\\
&&+
\frac{2 q.\sigma _2 \left(7 m_d+
            m_u\right) \left(\left(224 x^2-9\right) m_d+3 \left(224 x^2+19
\right) m_u\right) }{63 \left(28 x^2+3\right)^2 b_N^2 m_d^2
m_u^2}\sigma _3. \sigma _4~q.\sigma _2
\nonumber\\
&&
-\frac{8 \left(7 x^2+1\right)  \left(11
   m_d^2+\left(4 x^2+7\right) m_u m_d+2 \left(6
   x^2+1\right) m_u^2\right) \sigma _2 \sigma _3}{3
   \left(28 x^2+3\right)^2 b_N^2 m_d^2 m_u^2} \sigma _2 .\sigma _3~q.\sigma _4
\earr
where $x=\frac{b_{\pi}}{b_N}$. In the special case $m_u=m_d=\frac{m_N}{3}$ we get for the local terms:
\barr
{\bf A}_2.{\bf A}_3&=&
-\frac{64 \left(7 x^2+1\right) \left(56 x^2+3\right)^2 }{441 \left(28 x^2+3\right)^3 m_N^2}
q.\sigma _2 q.\sigma \
_3 q.\sigma _4
+\frac{256 \left(56 x^2+3\right) }{63 \left(28
x^2+3\right)^2 b_N^2 m_N^2}  \sigma _2.\sigma _4~q.\sigma _3
\nonumber\\
&&+
\frac{256 \left(56 x^2+3\right) }{63 \left(28
x^2+3\right)^2 b_N^2 m_N^2}  \sigma _3.\sigma _4~q.\sigma _2
 -\frac{32 \left(4 x^2 + 5\right) \left(7 x^2 +
          1\right) }{3 \left(28 x^2 +
            3\right)^2 b_N^2 m_N^2} \sigma _2. \sigma _3  ~q.\sigma _4
  \earr
 while the non local terms become:
\barr
{\bf A}_2.{\bf A}_3&=&
\frac{16 \left(56 x^2+3\right)^2 }{441 \left(28 x^2+3\right)^3 m_N^2}
p_N.\sigma _2 p_N.\sigma _3 p_N.\sigma_4
-\frac{256 \left(56 x^2+3\right) }{63
\left(28 x^2+3\right)^2 b_N^2 m_N^2} \sigma _2 .\sigma _4~p_N.\sigma _3
\nonumber\\
&&
-\frac{256 \left(56 x^2+3\right) }{63
\left(28 x^2+3\right)^2 b_N^2 m_N^2}  \sigma _3. \sigma _4~p_N.\sigma _2
+\frac{8 \left(4 x^2 +
          5\right) }{3 \left(28 x^2 +
            3\right)^2 b_N^2 m_N^2}  \sigma _2 .\sigma _3 p_N.\sigma _4
\earr
The first term of the local equation can be cast in the more suitable form by noting that:
\beq
\sigma _2.q \sigma _3.q \sigma _4.q=\frac{q^2}{3} \left ( \sigma _2. \sigma
_3 \sigma _4.q-\frac{2 \sqrt{3}}{\sqrt{5} }\left[  (\sigma _2\times \sigma
_3) k_{12}=2\times \sigma _4 \right]^{k=1}.q \right )
\eeq
Using the spin matrix elements of table \ref{table.spinme} we finally get using the current quark masses
\barr
{\bf A}_2.{\bf A}_3&=& 5 \sqrt{2} q^2
\nonumber\\
&&
[
\frac{\left(7 x^2+1\right) \left(21
   \left(14 x^2+1\right) m_d-\left(70 x^2+9\right)
   m_u\right) \left(\left(224 x^2-9\right) m_d+3
   \left(224 x^2+19\right) m_u\right)}{1323 \left(28
   x^2+3\right)^3 m_d^2 m_u^2}
   \nonumber\\
&&+26 \sqrt{2}
\nonumber\\
&&\frac{ \left(-7 \left(224 x^2-75\right)
   m_d^2+2 \left(1176 x^4-938 x^2+93\right) m_u
   m_d+\left(7056 x^4+2212 x^2+201\right)
   m_u^2\right)}{567 \left(28 x^2+3\right)^2 b_N^2 m_d^2
   m_u^2} ]
   \nonumber\\
   &&
\sigma_N.q
   \earr
   where $\sigma_N$ is the nucleon spin,
   while for the constituent quark masses we get
    \barr
{\bf A}_2.{\bf A}_3&=&
\left[q^2\frac{320 \sqrt{2}  \left(7 x^2+1\right) \left(56
   x^2+3\right)^2}{147 \left(28 x^2+3\right)^3 m_N^2}
+
\frac{416 \sqrt{2} \left(588 x^4-77 x^2+57\right)}{63
   \left(28 x^2+3\right)^2 b_N^2 m_N^2}\right]
   \nonumber\\
   &&
 \sigma_N.q
   \earr
   Note the presence of the $q^2$ in the first term. This will lead to an operator with a different
radial dependence, i.e.  $F^{(k)}_i(x)\Longrightarrow -\nabla ^2  F^{(k)}_i(x)$ (see Eq. (\ref{Eq:Frs})).
\end{enumerate}

  \end{document}